\documentclass[pra,superscriptaddress,twocolumn]{revtex4}
\usepackage[latin1]{inputenc}
\usepackage{amsmath}
\usepackage{amssymb}
\usepackage{dsfont}
\usepackage{color}
\makeatletter
\usepackage{amssymb}
\usepackage{graphicx}
\usepackage{epstopdf}
\usepackage{epsfig}
\usepackage{subfigure}
\makeatother

\usepackage{cancel,ifthen}
\usepackage[normalem]{ulem}

\newcommand{\mje}[1]{{\color{black}#1}}
\newcommand{\me}[1]{{\color{black}#1}}
\begin{document}

\title{Coherent Impurity Transport in \mje{an Attractive} Binary Bose-Einstein condensate}

\author{M. J. Edmonds}
\affiliation{Quantum Systems Unit, Okinawa Institute of Science and Technology Graduate University, Okinawa 904-0495, Japan}
\author{J. L. Helm}
\affiliation{Department of Physics, University of Otago, Dunedin, New Zealand}
\author{Th. Busch}
\affiliation{Quantum Systems Unit, Okinawa Institute of Science and Technology Graduate University, Okinawa 904-0495, Japan}

\date{\today{}}

\begin{abstract}\noindent
We study the dynamics of a soliton-impurity system modeled in terms of a binary Bose-Einstein condensate. This is achieved by `switching off' one of the two self-interaction scattering lengths, giving a two component system where the second component is trapped entirely by the presence of the first component. It is shown that this system possesses rich dynamics, including the identification of unusual `weak' dimers that appear close to the zero inter-component scattering length. It is further found that this system supports quasi-stable trimers in regimes where the equivalent single-component gas does not, which is attributed to the presence of the impurity atoms which can dynamically tunnel between the solitons, and maintain the required phase differences that support the trimer state. 
\end{abstract}
\maketitle
\section{Introduction}

\noindent Multi-component matter plays host to a plethora of novel phenomena, at both the classical and quantum mechanical level. The coexistence of several coupled, interacting degrees of freedom can facilitate different phases of matter, such as the miscible-immiscible phase-separation of binary fluids, arising from energetic competition between the differing components of the fluid \cite{cond_book}.

Quantum fluids - systems of interacting particles comprised of Fermions or Bosons cooled below their respective degeneracy temperature, can now be used to give direct insight into many analogous systems due to their high degree of experimental controllability. In particular, it is now feasible to engineer the dimensionality \cite{gorlitz_2001,gerbier_2004}, particle interactions \cite{chin_2010} and potential landscape \cite{henderson_2009} of these macroscopic systems. Complementary to this, the optical manipulation of these systems has reached maturity - oppurtunities now exist to emulate  complex phases of matter in the presence of gauge fields \cite{dalibard_2011,goldman_2014}, which form a key ingredient for many condensed matter effects of interest.

Solitary waves have been produced experimentally in both single and multi-component condensate systems. In the former case, quasi-stable soliton states have been generated, comprising single \cite{khaykovich_2002} as well as trains of bright solitons \cite{cornish_2006,strecker_2002}. Further work demonstrated bright solitons sensitivity to surface physics in the form of both repulsive \cite{marchant_2013} and attractive potentials \cite{marchant_2016}. Understanding the observed stability of these fragile systems has revealed the important role the complex phase of the matter-wave plays in these systems \cite{nguyen_2014,nguyen_2017}. Matter-wave solitons have been touted for applications in metrology, where these state's inherent coherence advocates them as strong candidates for engineering matter-wave interferometry \cite{martin_2012,helm_2015,helm_2018,haine_2018}. This in particular has led to the realisation of a matter-wave bright soliton Mach-Zehnder interferometer with a $^{85}$Rb condensate \cite{mcdonald_2017}, as well as proposals to controllably split solitons \cite{billam_2011}, and very recently schemes to realise bright soliton states with minimal noise have appeared \cite{edmonds_2018}. The purity of cold atom systems has also been exploited to gain insight into the role disorder plays for the dynamics of bright solitonic states in cold atomic gases \cite{lepoutre_2016,boisse_2017}.
\begin{figure}[b]
\includegraphics[width=0.75\columnwidth]{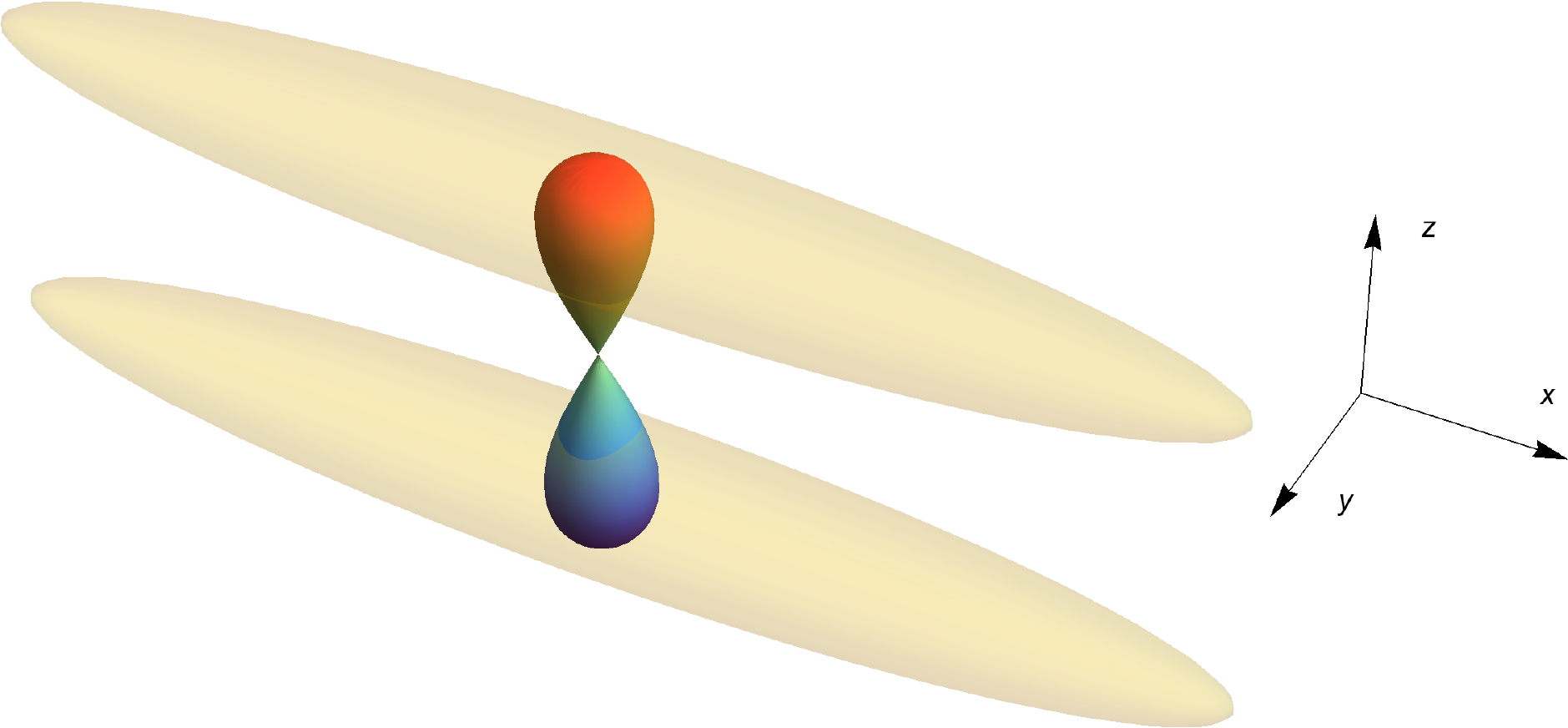}
\caption{\label{fig:share}{Schematic representation of the multi-soliton system. The two elongated condensate isosurfaces represent the three-dimensional density of the bright solitons, and the dumbbell shaped impurity is delocalized between both solitons}}
\end{figure}

There have also been experimental realizations of solitary wave structures in multi-component systems. Early theoretical work studied the properties of dark-bright and bright-bright solitons \cite{busch_2001,yang_2000} the first of which was  subsequently realized individually \cite{becker_2008} and also in the form of trains \cite{hammer_2011}. As well as this, studies have focussed on the role of potential barriers in the dynamics of vector solitons \cite{chang_2015}. Theoretical work has predicted that the single component focussing nonlinear Schr\"odinger equation can possess chaotic solutions in the presence of an axial harmonic potential \cite{martin_2007,martin_2008}, as well as the observed interaction induced frequency shift of pairs of trapped bright solitons \cite{martin_2016} in the experiment of Ref. \cite{nguyen_2014}. Complementary to this, theoretical work has focussed on solitary waves in higher spin systems, revealing the existance of integrable points in the full parameter space of the spin-1 condensate, in the form of so-called `polar' bright solitons \cite{ieda_2004,szankowski_2010}. Although solitons are usually studied as the solutions to one-dimensional nonlinear models, there have also been predictions of stable two-dimensional solitary wave solutions in dipolar Bose-Einstein condensates \cite{pedri_2005,tikhonenkov_2008}, where the additional nonlocal nonlinearity provides the stabilizing mechanism for these solitons. Very recently the Jones-Roberts soliton was realized experimentally, a true two-dimensional solitary wave structure \cite{myer_2017}.     

The realization of artificial electromagnetism with cold gases, and in particular spin-orbit coupling for Bose-Einstein condensates opens a novel route towards studying nonlinear wave structures. Here, the coupling of the condensates momentum to a quasi-spin leads to stripe-like soliton phases, related to the underlying immiscible phase of these systems \cite{xu_2013,achilleos_2013}. Spin-orbit coupling forms a key ingredient in simulating more exotic scenarios, such as Dirac-like equations, where confined solutions have been predicted \cite{merkl_2010} that resemble their bright soliton cousins in single component condensates.

Atomic condensates benefit from being exceptionally pure systems - this in turn allows one to investigate the effects of disorder and defects with an unprecedented level of control. 
The presence of impurities in ensembles of ultracold matter has led to predictions of impurity-molecules and lattices at the mean-field level \cite{li_2013}, as well as the role of many-body correlations for a single impurity out-of-equilibrium \cite{kronke_2015}. Experimental work has studied the role that spin impurities have in the strongly correlated Tonks-Girardeau limit \cite{palzer_2009} and also magnetic spin models \cite{fukuhara_2013}, which have also been the focus of subsequent theoretical investigations \cite{rutherford_2011,goold_2010,johnson_2015}. Complementary to this, recent experimental advances have led to the realization of trapping one matter-wave inside another, where a degenerate Fermi gas of $^{6}$Li atoms was confined inside a $^{133}$Cs Bose-Einstein condensate \cite{desalvo_2017}.

\mje{The ability to both prepare and control ultracold gas experiments gives access to physical regimes that mimic and go beyond those associated with conventional condensed matter physics. Impurities play a central role in condensed matter, since most materials will contain some imperfections. One important example drawn from this field is the \textit{polaron}, a quasi-particle that consists of an \me{electron} and the distortion caused by the passage of the \me{electron} through the ionic lattice. Impurities in the form of polarons can act as a sensitive probe within many-particle systems, and can be used to explore the correlations of these systems. \me{Additionally it should be noted that polarons are not necessarily dependent on the presence of impurities in a material, they can also appear in ideal crystals.} Over the last few years, ultracold gas experiments have succeeded in simulating the physics of polarons, including the pioneering experimental realisation of polarons of both bosonic \cite{jorgensen_2016,hu_2016} and fermionic \cite{schirotzek_2009} gases. The physics of polarons has also formed an ongoing focus of theoretical investigations. Optical lattices yield access to many models of interest in condensed matter physics, however as they are constructed from the interference of two counter-propagating laser modes, they do not naturally yield lattice vibrations (phonons), a key ingredient for polaron physics. This important question was investigated in \cite{bruderer_2007}, which proposed a methodology to overcome this drawback. Further work investigated the effect of dimensionality on the self-trapping of impurities, revealing regions where stable polarons can exist \cite{bruderer_2008}. Very recently, a theoretical investigation has revealed the universal behaviour of the bosonic polarons energy and its dependence on the Efimov parameter \cite{yoshida_2017}.}   

In this publication we will outline the collisional dynamics of multicomponent soliton-impurity systems, and how binary or triplet collisions might be exploited to perform deterministic population transfer operations on the impurity, providing a toolkit for future applications to metrology and quantum computation. The soliton-impurity system at the heart of our work is shown schematically in Fig.~\ref{fig:share}, where two soliton isosurfaces are shown with the delocalized impurity component. The paper is organized as follows. In Section \ref{sec:stab} we examine the stability of this system using a full three dimensional variational approach in order to understand the regimes where stable dynamics can be realized. Then in Section \ref{sec:model}, we state the model for the two component system in terms of coupled mean-field Gross-Pitaevksii equations for the dynamics. After this in Section \mje{\ref{sec:single} we explore the ground states of the binary system, following which in Sec.} \ref{sec:dyn} we undertake a scattering analysis of a single soliton molecule carrying an impurity with an `empty' soliton. We then proceed to show how soliton molecule complexes can be built using three solitons in Section \ref{sec:tri}, and study the resulting \mje{nonlinear} dynamics of the solitons and impurity as a function of the relative phase and inter-component scattering length, \mje{revealing the coherent nature of the impurities dynamics. We also discuss the conditions under which this state is stable to thermal fluctuations, before demonstrating that the impurity undergoes a novel localization transition.} We conclude with a summary of our findings in Section \ref{sec:sum}.   

\section{\label{sec:stab}Soliton-Impurity Stability}

\noindent The majority of experiments with atomic condensates are realized with repulsive inter-particle interactions confined by harmonic potentials. Under these conditions, the condensate is unconditionally stable. The introduction of attractive interactions can lead to a collapsed state, originating in the dispersive kinetic energy of the gas being overwhelmed by the attractive interactions between particles. We consider a two component model, where the second component of the system can be modeled as an `impurity', since the mass (number of atoms) of either component can be independently varied \cite{kalas_2006,kasamatsu_2006}. We consider a two component (binary) system forming a Bose-Einstein condensate coupled via completely attractive mean-field interactions. The stability of such a system depends on a number of parameters, in-particular the various scattering lengths, the number of atoms in each component and also the trapping geometry. To gain insight into the collapse dynamics of the binary system, consider the energy functional
\me{\begin{equation}\label{eqn:en3d}
E[\Psi_1,\Psi_2]=\int d^{3}{\bf r}\bigg[\sum_{j}H_{0j}+\sum_{j,k}\frac{g_{jk}}{2}|\Psi_{j}|^2|\Psi_{k}|^2\bigg],
\end{equation}}
where the wave function of component $j$ is $\Psi\equiv\Psi_j({\bf r})$, and the $s$-wave scattering length \me{$a_{jk}$} is contained in the parameter \me{$g_{jk}=4\pi\hbar^2a_{jk}/m$} where $m$ is the atomic mass. Note that in this system there are only two scattering \mje{parameters depending on the various scattering lengths}, $g_{11}$ and $g_{12}$, while $g_{22}=0$, and so the second component is linear and moves in the effective potential defined by the first component. The single-particle Hamiltonian $H_{0j}$ appearing in Eq.~\eqref{eqn:en3d} is defined as 
\begin{equation}\label{eqn:ham0}
H_{0j}=\frac{\hbar^2}{2m}|\nabla\Psi_j|^2+\frac{1}{2}m\omega_{\perp}^{2}{\bf r}_{\perp}^{2}|\Psi_j|^2
\end{equation}
where $\omega_{\perp}$ defines the transverse trapping frequency of the cloud, and ${\bf r}_{\perp}^{2}=y^2+z^2$ defines the radial coordinate. Then, this problem contains three length scales, two associated with the two scattering lengths, as well as one from the harmonic trapping term appearing in Eq.~\eqref{eqn:ham0}. \mje{The collapse instability for the cylindrically symmetric single-component gas has been studied previously, including the effect of additional axial confinement \cite{gammal_2001}.}  
To understand the nature of the collapse, we employ the cylindrically symmetric Gaussian variational ansatz
\begin{equation}\label{eqn:3dg}
\Psi_{j}=\sqrt{\frac{N_j}{\sqrt{\pi^3}\sigma_{xj}\sigma_{\perp j}^{2}\ell_{\rm ho}^{3}}}\exp\bigg(-\frac{1}{2\ell_{\rm ho}^{2}}\bigg[\frac{x^2}{\sigma_{xj}^{2}}+\frac{{\bf r}_{\perp}^{2}}{\sigma_{\perp j}^{2}}\bigg]\bigg)
\end{equation} 
Equation \eqref{eqn:3dg} introduces two pairs of dimensionless variational parameters, $\sigma_{xj}$ and $\sigma_{\perp j}$ which define the axial and transverse widths of the cloud respectively. The length scale $\ell_{\rm ho}=\sqrt{\hbar/m\omega_{\perp}}$ is defined using the transverse harmonic trapping frequency. Lastly, the normalization of each component is defined as $\int d^{3}{\bf r}|\Psi_j|^2=N_j$ where $N_j$ is the atom number in each component. Note that the ansatz of Eq.~\eqref{eqn:3dg} is appropriate since both scattering lengths are attractive, so the system is miscible with both components spatially overlapping. It is also possible to consider the immiscible case, where one scattering length is repulsive and the other attractive, which has also been shown to support stable solitary wave structures \cite{yakimenko_2012} in a quasi two-dimensional scenario. We can then insert the ansatz Eq.~\eqref{eqn:3dg} into Eq.~\eqref{eqn:en3d}, yielding
\me{\begin{align}\nonumber
&\frac{E}{\hbar\omega_{\rm ho}N_{1}}=\sum_{j}\frac{N_{j}}{N_1}\bigg[\frac{1}{4\sigma_{xj}^{2}}+\frac{1}{2\sigma_{\perp j}^{2}}+\frac{1}{2}\omega_{\perp}^{2}\sigma_{\perp j}^{2}\bigg]\\&{+}\frac{1}{\sqrt{2\pi}\ell_{\rm ho}}\frac{a_{\rm 11}N_1}{\sigma_{x1}\sigma_{\perp 1}^2}{+}\frac{1}{\sqrt{\pi}\ell_{\rm ho}}\frac{4a_{\rm 12}N_2}{\sigma_{\perp 1}^{2}+\sigma_{\perp 2}^{2}}\frac{1}{\sqrt{\sigma_{x1}^{2}+\sigma_{x2}^{2}}},
\label{eqn:var3den}
\end{align}}
where Eq.~\eqref{eqn:var3den} introduces $N_2/N_1$ as the mass imbalance. Then the collapse point of the system can be found for a particular set of parameters by simultaneously solving the pair of equations \cite{carr_2002}
\begin{equation}\label{eqn:pair}
\nabla E=\underline{0}\hspace{0.5cm}\text{and}\hspace{0.5cm}\text{det}(\mathcal{J}\nabla E)=0,
\end{equation}
where $\nabla=\sum_j ( \hat{e}_{xj}\partial_{\sigma_{xj}}+\hat{e}_{\perp j}\partial_{\sigma_{\perp j}} ) $ defines the four component gradient operator in the variational problem, and $\mathcal{J}$ is the associated Jacobian. Under general conditions, Eqs.~\eqref{eqn:pair} must be solved numerically to obtain the collapse point of the condensate for a given set of parameters. 
\begin{figure}[t]
\includegraphics[scale=0.75]{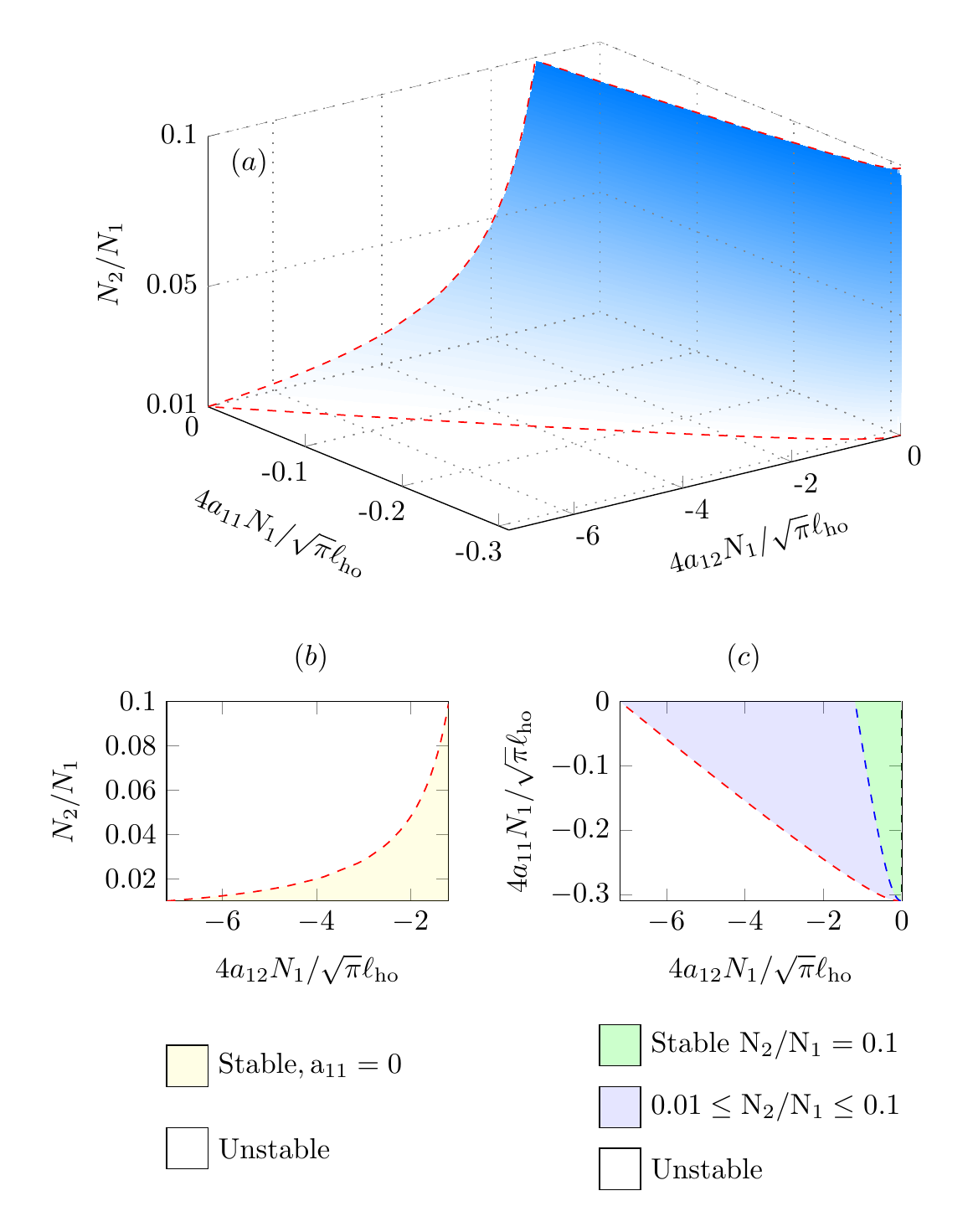}
\caption{\label{fig:coll}Stability of the Soliton-Impurity system in the $a_{11}$-$a_{12}$-$N_j$ parameter space, (a). The red lines indicate boundaries to collapse in different planes. The lower panels (b) and (c) show cross-sections of the data presented in (a). The volume enclosed by the surface contains the parameter space of the model that is stable to mean-field collapse.}
\end{figure}
Figure \ref{fig:coll} shows the numerically obtained solutions to Eqs.~\eqref{eqn:pair}. These solutions are obtained using an iterative procedure to procure the collapse point starting from a point in the parameter space with known analytical solution, in this case the point $g_{\rm 12}=0$, from which the critical collapse point for a cylindrically symmetric trap is $N\me{a_s}/a_{\perp}=-1/\sqrt[4]{3}$\mje{, where $N$ is the atom number and $a_s$ the $s$-wave scattering length.} The mean-field collapse phase diagram is shown in Fig.~\ref{fig:coll}(a), the volume enclosed by the $a_{11}$, $a_{12}$ and $N_2/N_1$ axis define the space of stable three dimensional solitons. Here the red lines show the boundary between stable and unstable regimes in each parameter plane. It can be seen that when $N_2/N_1$ is small, corresponding to a small impurity population the collapse point is moved to larger values of \me{$a_{\rm 12}$}. As the number of atoms in the impurity $N_2$ increases, the collapse point in the \me{$a_{\rm 12}$-$a_{11}$} plane moves to smaller values of $a_{\rm 12}$. This result is intuitive, since one can interpret the additional attractive inter-species mean-field potential as providing an extra destabilizing contribution to the mean-field energy.

Fig.~\ref{fig:coll}(b) and (c) show cross-sections of the parameter space presented in Fig.~\ref{fig:coll} (a). Panel (a) shows a cut through the plane $a_{\rm 11}=0$, where the stable (white) and unstable (yellow) region are separated by the dashed red line. The second panel, (b) shows a different cut through (a) for constant $N_2/N_1$. The green shaded region bounded by the blue dashed line indicates the stable region for $N_2/N_1=0.1$ in the \me{$a_{\rm 11}$-$a_{\rm 12}$} plane. The blue shaded region is stable for $N_2/N_1=0.01$, but not $N_2/N_1=0.1$. Again, the white region is unstable to collapse. This rudimentary analysis shows that any experiment to realize a fully attractive two-component system would be favorable to a moderate mass imbalance, especially if one was interested in exploring the dynamics as a function of one of the scattering lengths of this system, as we will proceed to do in the following sections of this work. \mje{Since we consider a mean-field mass-imbalanced system it is worth considering when such a model is valid. It is known for example that on the repulsive side ($\me{a_{jk}}>0$), this model undergoes composite fermionization \cite{zollner_2008}. We expect this attractive mean-field model to be suitable up to the collapse point, although it is conceivable that fluctuations could play an important role in this mass imbalanced system. However, this analysis lies beyond the scope of the current work.}   

\section{\label{sec:model}Equations of motion}

\noindent One of the characteristic attributes of solitary waves are their particle-like properties \cite{drazin_1989}. Consequentially, their inherent robustness leads to collision dynamics where they emerge unscathed, with the exception of a phase shift. For the single-component focussing nonlinear Schr\"odinger equation, the scattering of two bright solitons is always elastic, a consequence of the underlying integrability of the nonlinear Schr\"odinger equation. For the two component system the equations of motion for $\Psi_{j}({\bf r},t)$ are found from the Lagrangian density 
\me{\begin{align}\nonumber
\mathcal{L}({\bf r},t)&=\hbar\sum_{j}\text{Im}\bigg[\partial_t\Psi_{j}^{*}({\bf r},t)\Psi_{j}({\bf r},t)\bigg]\\&-\bigg\{\sum_{j}H_{0j}+\sum_{j,k}\frac{g_{jk}}{2}|\Psi_j|^2|\Psi_k|^2\bigg\},\label{eqn:lag}
\end{align}}
and the associated Euler-Lagrange equations. We are interested in studying the soliton solutions which exist in the quasi one-dimensional limit. As such, we assume that there is tight radial confinement, such that any radial dynamics are effectively frozen out. Then the radial dynamics for both components can be factorized in the form $\Psi_j({\bf r},t)=\psi_{j}^{\perp}({\bf r}_\perp)\psi_j(x,t)$ where $\psi_{j}^{\perp}({\bf r}_{\perp})=(1/\ell_{\perp}\sqrt{\pi})\exp(-{\bf r}_{\perp}^{2}/2\ell_{\perp}^{2})$ defines the ground state of the radial trap. Proceeding, the dynamics in the quasi one-dimensional limit are captured by
\me{\begin{subequations}\label{eqn:gpe1d}
\begin{align}
i\hbar\frac{\partial\psi_1}{\partial t}{=}&\bigg[{-}\frac{\hbar^2}{2m}\frac{\partial^2}{\partial x^2}{+}\frac{g_{11}}{2\pi \ell_{\perp}^{2}}|\psi_1|^2{+}\frac{g_{12}}{2\pi \ell_{\perp}^{2}}|\psi_2|^2\bigg]\psi_1,\\
i\hbar\frac{\partial\psi_2}{\partial t}{=}&\bigg[{-}\frac{\hbar^2}{2m}\frac{\partial^2}{\partial x^2}{+}\frac{g_{12}}{2\pi \ell_{\perp}^{2}}|\psi_1|^2\bigg]\psi_2,
\end{align}
\end{subequations}
with normalization
\begin{equation}
\int dx |\psi_{j}(x,t)|^2 = N_j.
\end{equation}}
The equations of motion defined by Eq.~\eqref{eqn:gpe1d} will form the work-horse for studying the binary attractive system. \me{This mean-field model was originally studied by Ref.~\cite{sacha_2006} who analyzed the localized solutions and their quantum fluctuations.} We note that this model has also been studied recently in the context of repulsive mean-field interactions, where it was shown how the dark soliton solutions long lifetimes can be used to host qubits for quantum information applications \cite{shaukat_2017}. \me{Complementary to this the physics of polarons remains a topic of ongoing interest, with very recent theoretical work focussing on studying so-called \textit{Fr\"olich} polarons \cite{grusdt_2017}; as well as the binding properties of trapped bosonic polarons \cite{dehkharghani_2018}.}
\begin{figure}[t]
\includegraphics[scale=0.8]{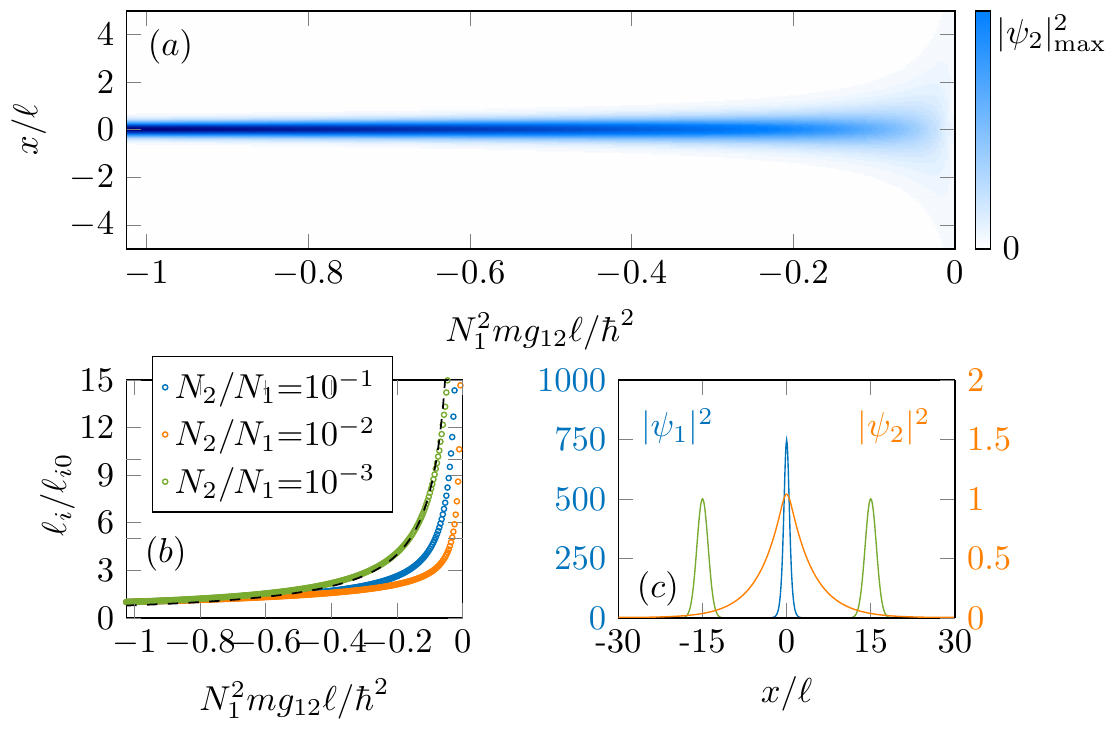}
\mje{\caption{\label{fig:singleimp}Single polaron ground sates. (a) shows the ground states of  Eqs.~\eqref{eqn:gpe1d} calculated as a function of the inter-component scattering \mje{parameter} \me{$g_{\rm 12}$}, while (b) shows the impurity length scale $\ell_i=\sqrt{\langle x^2\rangle}$ (size) computed again as a function of \me{$g_{\rm 12}$}. \me{The black dashed line shows a comparison with $\ell_{\rm i}=0.8\hbar^2/(N_{1}^{2}m|g_{\rm 12}|\ell)$}. Panel (c) shows an example ground state for $N_{1}^{2}m\me{g_{\rm 12}}\ell/\hbar^2=-1/8$, with $N_2=10$ impurity atoms. The green data shows an example of the initial multi-soliton states used for the binary and trimer simulations studied in Sec. \ref{sec:tri}.}}
\end{figure}
\begin{figure*}[t]
\includegraphics[width=2\columnwidth]{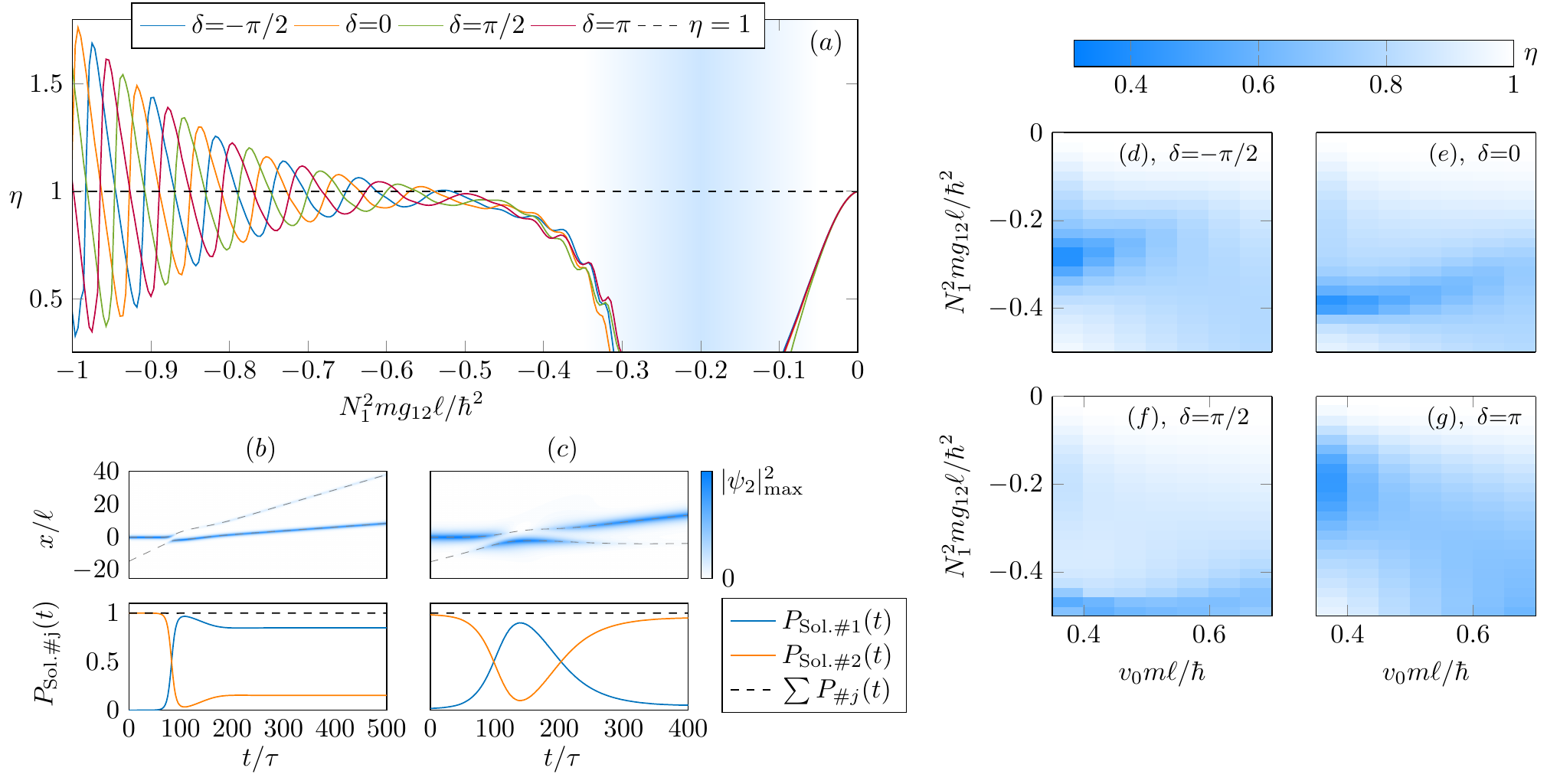}
\caption{\label{fig:eta}Soliton-Impurity dynamics. (a) shows the coefficient of restitution, Eq.~\eqref{eqn:eta} computed as a function of the inter-component scattering \mje{parameter}, $\me{g_{\rm 12}}$ for four equally spaced initial phase differences. (b) and (c) show example \mje{in-phase ($\delta{=}0$)} dynamics for $N_{\rm 1}m\me{g_{\rm 12}}\ell/\hbar^2\simeq-1.01/N_1$ (left column) and $N_{\rm 1}m\me{g_{\rm 12}}\ell/\hbar^2\simeq-0.24/N_1$ (right column). The top row in each case shows the space-time propagation of the impurity, with the trajectories of the soliton component plotted as a gray dotted line. The bottom row show the respective population of the impurity in each soliton as a function of time. (d)-(g) shows $\eta$ calculated as a function of $N_{\rm 1}m\me{g_{\rm 12}}\ell/\hbar^2$ and $mv_0\ell/\hbar$, the (dimensionless) initial velocity of the (empty) soliton, again for four equally spaced phase differences, $\delta=-\pi/2,0,\pi/2,\pi$.}
\end{figure*}
\mje{
\subsection{\label{sec:single}Single Polaron Ground states}
To understand the basic physics of the attractive binary condensate defined by Eqs.~\eqref{eqn:gpe1d}, we begin by computing the ground state of this system as a function of the inter-component scattering \mje{parameter} \me{$g_{\rm 12}$}. This is shown in Fig.~\ref{fig:singleimp} (a), which shows the density $|\psi_2|^2$ of the impurity for $N_1=10^3$, $N_2=10$. Here $N_{1}^{2}m\me{g_{\rm 11}}\ell/\hbar^2=-6$. For large negative values of the inter-component scattering length, the impurity is well localized within the soliton. As $\me{g_{\rm 12}}\rightarrow0$, the width of the impurity wave function starts to grow. This effect is investigated further in panel (b), where we compute the effective width (standard deviation) of the impurity, $\ell_i = \sqrt{\langle x^2\rangle}$ as a function of $\me{g_{\rm 12}}$, for different mass ratios $N_2/N_1$. Each individual data set is scaled to the width of the impurity for the largest negative scattering length ($\ell_{\rm i0}$) for ease of comparison. One can see that as $\me{g_{\rm 12}}\rightarrow0$, this quantity increases by an order of magnitude from its smallest value. \me{The black dashed line shows a comparison of $\ell_{\rm i}/\ell_{\rm i0}$ for $N_2/N_1=10^{-3}$ with the power-law $\ell_{\rm i}=0.8\hbar^2/(N_{1}^{2}m|g_{\rm 12}|\ell)$, where the numerical value $0.8$ is a fitting parameter. For other mass ratios this agreement breaks down, due to the increased influence of the impurity component on the overall shape of the soliton.} The final panel of Fig.~\ref{fig:singleimp}, (c) shows an example ground state. Here $N_{1}^{2}m\me{g_{\rm 12}}\ell/\hbar^2=-1/8$, the soliton is shown in blue ($|\psi_1|^2$), while the orange data is the impurity ($|\psi_2|^2$). Also included in green are the extra solitons that are presented later in the first component for the trimer simulations in Sec.~\ref{sec:tri}. 
}
\subsection{\label{sec:dyn}Binary Soliton-Impurity Dynamics}
\noindent To gain insight into the dynamics of the soliton impurity system, we simulate collisions between an `empty' bright soliton, that is a soliton solution obtained for $g_{12}=0$ from Eq.~\eqref{eqn:gpe1d} with the soliton containing an impurity. As such, our initial condition takes the form
\begin{equation}\label{eqn:psi1d0}
\Psi_{\rm 2Sol}(x,v)=\bigg(\begin{array}{c}\psi_1 \\ \psi_2\end{array}\bigg)+\bigg(\begin{array}{c}1 \\ 0\end{array}\bigg)\psi_{\rm S}(x-x_0,v),
\end{equation}
where $\psi_j$ represents the numerical solution to Eq.~\eqref{eqn:gpe1d} for a given finite choice of $g_{11}$ and $g_{12}$. \mje{The first component $\psi_1$ contains the first bright soliton which hosts the impurity ($\psi_2$) localized at the origin $x=0$. The} length scale (size) of each component depends critically on the scattering parameters \me{$g_{\rm 11}$} and \me{$g_{\rm 12}$}. The function $\psi_{\rm S}(x)$ is the single soliton solution given by
\begin{equation}\label{eqn:1dsol}
\psi_{\rm S}(x,v)=\sqrt{\frac{N_1}{4\ell_1}}\text{sech}\bigg(\frac{x}{2\ell_1}\bigg)\exp\bigg(i\frac{mv}{\hbar}[x+x_0]+i\delta\bigg)
\end{equation}
and the length scale appearing in Eq.~\eqref{eqn:1dsol} is $\ell_{\rm 1}=\hbar^2/(m|g_{\rm 11}|N_{\rm 1})$, with $N_{\rm 1}$ giving the number of atoms in each soliton of the first component $\psi_1$, \mje{ while the scaled quasi-one-dimensional scattering \mje{parameter} is} $g_{\mathrm{11}}=2\hbar a_{11}\omega_{\perp}$ \mje{ which describes} the solitonic nonlinearity strength. The initial phase difference is given by $\delta$. The parameter space associated with Eq.~\eqref{eqn:gpe1d} contains two scattering lengths, two atom numbers, the initial velocity $v_0$ and position $x_0$ of the soliton and impurities as well as the initial phase difference, and as such is generally complicated to understand completely. To draw out the main features of the model, we simulate collisions for fixed $g_{11}$ and atom number, but vary the inter-component scattering length $g_{12}$.

A useful measure for soliton collisions with non-integrable dynamics is the coefficient of restitution. This is a dimensionless quantity defined as the total kinetic energy of two particles after a collision to the total kinetic energy before the collision \cite{edmonds_2017,dingwall_2018}
\begin{equation}\label{eqn:eta}
\eta=\frac{(m_{1}v_{1}^{2}+m_{2}v_{2}^{2})_{f}}{(m_{1}v_{1}^{2}+m_{2}v_{2}^{2})_{i}}.
\end{equation}
Now, if $\eta=1$ the collision is elastic with conserved momenta before and after collisions, while $\eta\neq1$ indicates an inelastic collision between the solitons. The masses $m_i$ and velocities $v_i$ appearing in Eq.~\eqref{eqn:eta} are computed from
\begin{subequations}\label{eqn:mandv}
\begin{align}
m_{j}&=m\int dx|\psi_{1}(x)|^2,\\
v_{j}&=-\frac{i\hbar}{m_{j}}\int dx \psi^{*}_{1}(x) \frac{\partial\psi_{1}(x)}{\partial x}.
\end{align}
\end{subequations}
Both quantities appearing in Eq.~\eqref{eqn:mandv} are computed {\it locally} around the center of mass of each individual soliton.

In our simulations presented in Fig.~\ref{fig:eta} we have taken $N_{\rm 1}m\me{g_{\rm 11}}\ell/\hbar^2=-6/N_1$, $mv_0\ell/\hbar=0.15$ is the dimensionless initial velocity, while $x_0=-15\ell$ is the initial displacement of the empty soliton. The normalization of both components is $\int dx|\psi_j|^2=1$, while each simulated collision is run for $t=500\tau$ units of real time. The numerical simulations are handled using a spectral (split-operator) method, and we work in the so-called soliton units \cite{martin_2008}, where $\ell=\hbar/mv$, $\tau=\ell/v$ and $E=mv^2$ define the units of length, time and energy respectively. \mje{To understand how these units correspond to physical quantities, we can use the experimental parameters of Ref.~\cite{marchant_2013}, who produced a bright solitary wave with a $^{85}$Rb condensate. Then one has $m=85u$, where $u$ is the atomic mass unit, $N_1=2,000$ atoms in each soliton, and a transverse trapping frequency of $\omega_\perp=27$Hz. Using these parameters one finds a natural length scale $\ell\simeq11\mu m$, and small value of $N_{\rm 1}^{2}m\me{g_{\rm 12}}\ell/\hbar^2\simeq-10^{-4}$, so in reality it would be necessary to use the powerful tool of Feshbach resonances in an experiment in order to bring the system into the regime described in this work.  }
\begin{figure}[b]           
\includegraphics[width=\columnwidth]{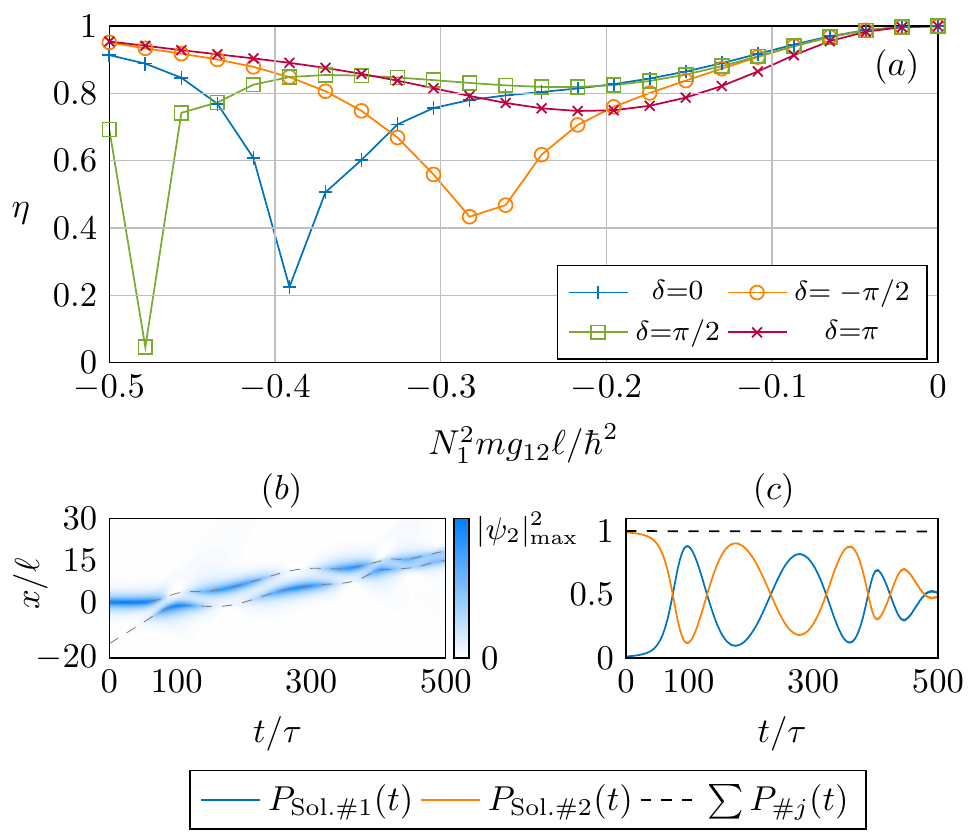}
\caption{\label{fig:dimer} Restitution data for fixed initial velocity $v_0m\ell/\hbar=0.4$ for different initial phase differences, (a). The two lower panels show example space-time dynamics for a `weak' dimer in (b) as well as the associated impurity populations of each soliton, $P_{\rm Sol.\#j}(t)$ in (c).}
\end{figure}
In figure \ref{fig:eta} we explore the binary dynamics of the Soliton-Impurity system. Here, a single empty soliton collides with the Soliton-Impurity system which is positioned initially at the origin. The coefficient of restitution, Eq.~\eqref{eqn:eta} is then computed as a function of the inter-component scattering length, $\me{g_{\rm 12}}$ for several initial phase differences, $\delta$. It should be noted that although the phase difference is set initially in our simulations, this quantity evolves dynamically \cite{khawaja_2011}, so the initial value is not necessarily the phase difference at the point of collision. This phase evolution can be inferred from $\eta$ as displayed in Fig.~\ref{fig:eta}(a), where we plot $\eta$, as a function of $\me{g_{\rm 12}}$, resulting from initial phase-differences $\delta=-\pi/2,0,\pi/2,\pi$. The dynamics of $\eta$ can be roughly partitioned into two regimes, a `weak' dimer phase (light-blue shading) that manifests for $0>N_{\rm 1}^{2}m\me{g_{\rm 12}}\ell/\hbar^2\gtrsim-0.5$, and a second more conventional non-integrable regime with $N_{\rm 1}^{2}m\me{g_{\rm 12}}\ell/\hbar^2\lesssim-0.5$. In the non-integrable regime the dependence of $\eta$ on the solitons' relative phases is demonstrated by the phase-winding of the $\eta$ oscillations associated with the different $\delta$s. Furthermore, $\eta$ is seen to oscillate with an increasing amplitude and frequency as $g_{\rm 12}$'s magnitude is increased. The frequency increases because increasing $\me{g_{\rm 12}}$ increases the chemical potential of the carrier soliton, and so its phase winds more quickly, effecting an additional phase-shift. \mje{Due to the non-integrability of this system, energy is not conserved if $\eta\neq1$. The energy of the solitons is redistributed post-collision. The primary mechanism for this is the change of the solitons masses post-collision. As well as this, some of the kinetic energy initially carried by the moving soliton is redistributed into potential energy post-collision, affecting an additional perturbation to the systems two scattering lengths.} In the `weak' dimer phase we note that all four curves meet at $g_{12}=0$ when $\eta=1$, where integrability is restored and the system is reduced to the single component focussing cubic Schr\"odinger equation. 

Example dynamics for each regime are displayed in Fig.~\ref{fig:eta}(b) and (c). The top left panel of (b) shows the space-time density $|\psi_{2}(x,t)|^2$ of the impurity for $N_{1}m\me{g_{\rm 12}}\ell/\hbar^2=\simeq-1.01/N_1$. The trajectories of the solitons are overlaid (gray dotted line). The lower left panel of Fig.~\ref{fig:eta}(b) shows the population of each soliton as a function of time
\begin{equation}\label{eqn:popi}
P_{\rm Sol. \#j}(t)=\int dx|\psi_2(x,t)|^2,
\end{equation}
which are calculated by integrating the density of the impurity locally around each solitons centre of mass. Then, for a system with $n_{\rm sol}$ solitons the total impurity population is given by
\begin{equation}
\sum_{j=1}^{n_{\rm sol}}P_{\rm Sol.\#j}(t)=N_2.
\end{equation}
The panels on the right column Fig.~\ref{fig:eta} (c) show the equivalent dynamics but for $N_1m\me{g_{\rm 12}}\ell/\hbar^2\simeq-0.24/N_1$. Clearly the dynamics of this system are highly non-integrable, showing a number of unusual dynamical effects. In particular, the second impurity component does not behave like a soliton, instead behaving like a quantum particle trapped by the potential generated by the first, solitonic component. Then, by exploring the scattering dynamics as a function of the inter-component scattering parameter $\me{g_{\rm 12}}$, per Fig.\ref{fig:eta}(a) one can interpret the dynamics of the system. For larger negative values of $\me{g_{\rm 12}}$, the impurity is localized deep within the soliton, and as such its length scale is typically less than that of the length scale (size) of the soliton in which it is initially localized. On the other hand, for smaller negative values of $\me{g_{\rm 12}}$ the potential felt by the impurity component is quite shallow, and the impurity in this regime is comparatively more weakly bound, having a length scale which can be significantly larger than that of its solitonic host. 

Since the impurity component feels the solitonic component as an effective dynamical potential, its dynamics can show some unusual features. In particular, the impurity can transfer itself into the other, empty soliton. This is shown in the lower panel of Fig.~\ref{fig:eta}(b), where the impurity population of each soliton is computed as a function of time. During dynamical evolution, $\sim$85\% of the initial impurity population is smoothly transferred from the second to the first soliton. In the second example shown in Fig.~\ref{fig:eta}(c) the initial population of the second soliton is almost completely transferred to the first, and then back again. \mje{We attribute the population transfer effect to quantum mechanical tunneling.} It is also worth mentioning that there is an additional subtletly in the interpretation of these dynamics. In the integrable limit, each soliton can be identified by its amplitude and velocity \cite{gordon_1983}, which means that the transfer of population between the two solitons in Fig.~\ref{fig:eta}(b) could also be interpreted with the labels of the two solitons switched, post collision. We will instead keep the labeling of the solitons more in the style of two potentials that the second impurity potential feels. This choice makes a quantitative but not qualitative difference to the interpretation of our results. These dynamics could be useful for atomtronics applications \cite{seaman_2007}, indeed these examples shown in Fig.~\ref{fig:eta}(b) and (c) behave somewhat like an analogue of a conventional transistor, where the first soliton that initially hosts the impurity can be interpreted as the `source' while the second empty soliton can be labelled the `drain', while the effective gate voltage is controlled by the inter-component scattering parameter, $\me{g_{\rm 12}}$.

Figure \ref{fig:eta}(d)-(g) explores the dynamics in the regime $-0.5\leq N_{\rm 1}^{2}m\me{g_{\rm 12}}\ell/\hbar^2\leq 0$, as a function of the initial velocity of the empty soliton, $v_0m\ell/\hbar$. For large initial velocities, the scattering is comparatively less sensitive to the scattering length $\me{g_{\rm 12}}$. However, as the initial velocity is lowered, a prominent dip develops, whose depth and position on the $\me{g_{\rm 12}}$ axis depends on the initial phase difference $\delta$. This effect is demonstrated in Fig.~\ref{fig:dimer}, which shows the data displayed from Fig.~\ref{fig:eta} reshaped. Each curve is taken for the fixed velocity $v_0m\ell/\hbar=0.4$. Here, one can see that the position and depth of the `dip' is quite sensitive to the initial phase difference, and seems to deepen for smaller scattering lengths as the phase difference is modulated. The lower row of panels in this figure show a clear example of the dimer state, for the choice of parameters $mv_0\ell/\hbar=0.1$, $\delta=0$ and $N_{\rm 1}m\me{g_{\rm 12}}\ell/\hbar^2\simeq-0.283/N_1$. Here, one can see the space-time evolution of the impurity in Fig.~\ref{fig:dimer} (b), with the dotted lines indicating the trajectories of the soliton in the first component. The impurities dynamics bare some resemblance to a braid, with the initially localized impurity oscillating around the centre of mass of the weak dimer. 

The second panel, Fig.~\ref{fig:dimer} (c) shows the impurity population $P_{\rm Sol.\#j}(t)$ of each soliton as a function of time. The solitons propagate together for quite sometime, with an oscillating impurity population. Contrasting the restitution data shown in Fig.~\ref{fig:dimer}(a) with that of Fig.~\ref{fig:eta}(a) is suggestive that the $\eta$ data actually merges for smaller values of the initial velocity, and only separates for larger values of $v_0$.  
\begin{figure*}[t!]
\makebox[\textwidth]{\includegraphics[width=2\columnwidth]{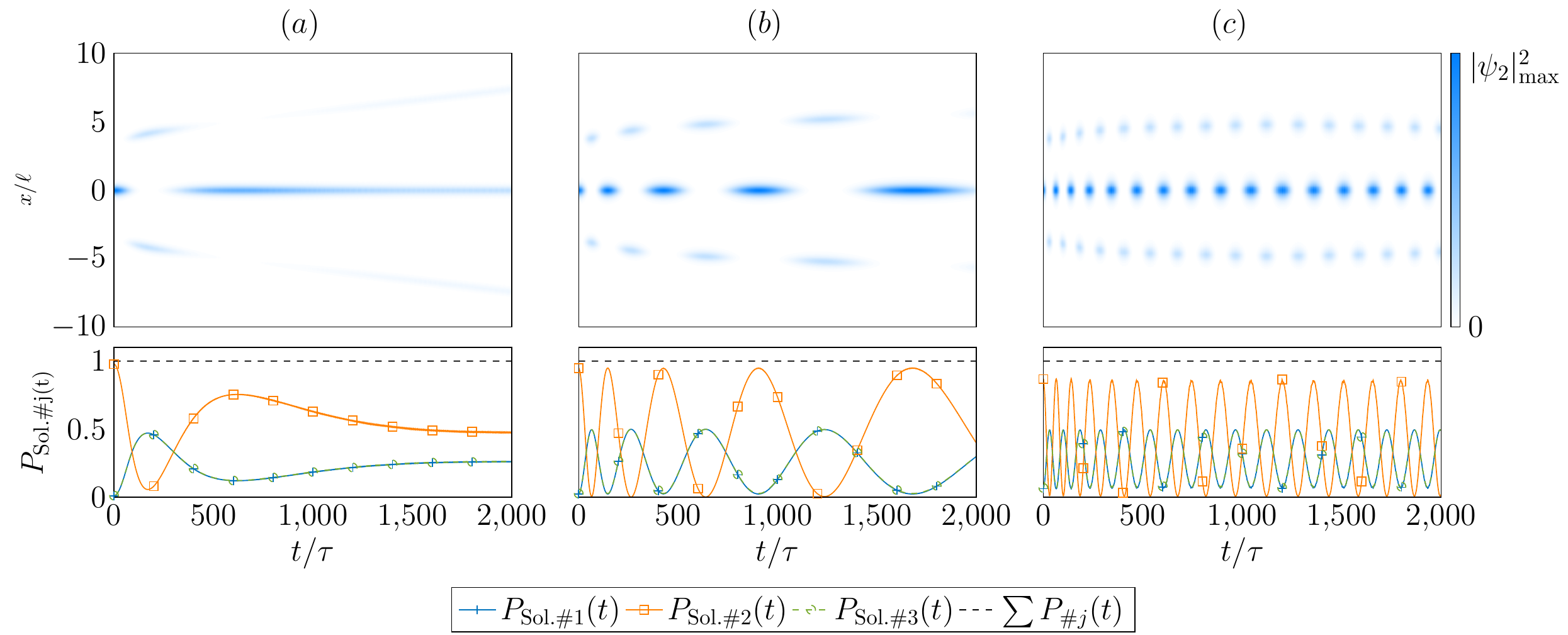}}
\caption{\label{fig:mol}Soliton trimer formation. From left to right, the inter-component scattering length is $N_{\rm 1}m\me{g_{\rm 12}}\ell/\hbar^2\simeq-(2.2,1.6,1)/N_1$ for (a), (b) and (c) respectively, and $N_{1}m\me{g_{\rm 11}}\ell/\hbar^2=-8/N_1$. The initial phase of the soliton positioned at $x=0$ is taken as $\delta=\pi/2$, while each soliton in the first component has $N_1=1,000$ atoms. The top row shows the space-time propagation of the impurity, $|\psi_2|^2$, while the lower panels show the impurity population of each soliton.}
\end{figure*}

The relative amount of kinetic to (attractive) potential energy in this system is crucial to the observed dynamics. Indeed, for collisions approaching zero inter-component scattering length, one has a highly delocalized impurity, which is weakly bound in its solitonic host. Coupled to this is the fact that the collision of the solitons in this system expels radiation in the form of small amounts of atomic density of the cloud. This can in turn interact with the solitons in this regime to further destabilize the observed dynamics. For scattering lengths slightly smaller in magnitude than Fig.~\ref{fig:dimer} (b) and (c), the post collision dynamics are found to be exceptionally sensitive to this radiation. This can partially be overcome by simulating collisions with increasingly larger numerical boxes, (in our work we typically use $L_{\rm box}=200\ell$) however as $|a_{\rm 12}|\rightarrow0$ the size of the impurity will always be larger than one can realistically simulate, an unavoidable limitation inherent to this system.

\section{\label{sec:tri}Soliton Molecules}
\subsection{\label{sec:mol}Soliton Trimers}
\noindent Models of nonlinear systems can also play host to higher-order soliton states, in the form of soliton molecules (i.e. several individual solitons forming bound objects) and also breathers, which are single solitonic entities that can be thought of as excited states of the focussing nonlinear Schr\"odinger equation, which have recently been engineered experimentally with matter waves for the first time \cite{everitt_2015} using an attractive gas of $^{85}$Rb. Soliton molecules have been studied in various guises within the context of ultracold matter, for example the realization of degeneracy with atomic species possessing significant dipole-dipole interactions has led to the prediction of novel molecular states in these systems \cite{baizakov_2015,bland_2015,pawlowski_2015}. Related to this are the realisation of `droplets' of both dipolar matter \cite{barbut_2016,schmitt_2016,baillie_2016} and intriguingly, also light with a non-trivial angular momentum structure \cite{wilson_2018}\mje{, as well as the prediction of soliton molecules in systems with nonlocal interactions \cite{salerno_2018}.}

Here, we consider a stationary spatially symmetric initial state to study the possibility of molecule-like states in the system described by Eqs.~\eqref{eqn:gpe1d}. In the previous section it was found that the low velocity scattering of a pair of solitons leads to increasingly inelastic dynamics as the initial kinetic energy in the system approaches zero. In fact, if we try to form a simple soliton molecule with a pair of initially stationary solitons, one with an impurity, and one without the resulting molecular state rapidly destabilizes. This is due to the impurity that causes the phase of the soliton in the first component to wind, eventually breaking the molecule. Instead we focus on understanding molecules formed from three individual solitons. In order to create an initially symmetric state, we must place the impurity either in the center soliton with the outer two solitons initially empty, or visa-versa. In our simulations we have chosen the former, so that the initial state is
\begin{equation}\label{eqn:itri}
\Psi_{\rm 3Sol}(x)=\bigg(\begin{array}{c}\psi_1 \\ \psi_2\end{array}\bigg)+\bigg(\begin{array}{c}1 \\ 0\end{array}\bigg)\sum_{j=1}^{2}\psi_{\rm S}(x-x_j,v_j{=}0),
\end{equation}
and $x_{j}$ are the centres of mass of the two outer solitons, $\psi_{\rm S}(x)$ is defined per Eq.~\eqref{eqn:1dsol}, and the $x_j$ are chosen symmetrically such that $x_1+x_2=0$. \mje{This initial configuration, built from the ground state and known exact solutions in the limit $g_{\rm 11}=0$ is also shown in Fig.~\ref{fig:singleimp} (c).} Figure \ref{fig:mol} shows example dynamics of the three soliton system. From left to right, panels (a) to (c) show long-time dynamics in the form of space-time density plots of the impurity $|\psi_2|^2$ (top row) while the bottom row shows the impurity population of each soliton as a function of time. Note that there are three curves in these figures, however the populations of the outer solitons are symmetric, so $P_{\rm Sol.\#1}$ and $P_{\rm Sol.\#3}$ are the same. The parameters used for the simulations here are $N_1m\me{g_{\rm 12}}\ell/\hbar^2\simeq-(2.2,1.6,1)/N_1$ corresponding to (a), (b) and (c) respectively. The initial phase of the central soliton is $\delta=\pi/2$. As the inter-component scattering length $\me{g_{\rm 12}}$ is increased, the dynamics of the system change quite drastically. This is reflected in that fact that in (a) the solitons move apart, with only one `switch' of population occurring during the dynamics, as shown in the lower panel of (a). As $\me{g_{\rm 12}}$ is increased, the impurity is delocalized, promoting tunneling to the outer solitons, as shown in (b). Finally in (c) a molecular-like state is formed, with the outer solitons showing a clear attraction towards the central soliton. The lower panel of (c) reflects this, where almost periodic oscillations of the impurity density are shown. 

\mje{\subsection{Thermal Fluctuations}

Given the fragile nature of bright soliton states, it is important to understand when the predicted soliton trimer presented in Fig.~\ref{fig:mol} is stable to thermal fluctuations that are present in real systems. One way to understand the conditions under which the trimer is stable to thermal fluctuations is to compare the energy difference between the absolute ground state of the system and the trimer state with the thermal energy present in the system. We denote each of these quantities by $E_{\rm gnd}$ and $E_{\rm tri}$ respectively. Then the energy difference $\delta E = E_{\rm tri} - E_{\rm gnd}$ we are interested in is given by 
\begin{align}\nonumber
\delta E &= \bigg(\sum_{j=1}^{3}E_{\rm kin}^{{\rm Sol.}j}+E_{\rm kin}^{\rm I}+\sum_{j=1}^{3}E_{\rm vdW}^{{\rm Sol.}j}+E_{vdW}^{\rm SI}\bigg)\\&-\bigg(E_{\rm kin}^{{\rm Sol.}2}+E_{\rm kin}^{\rm I}+E_{\rm vdW}^{\rm S2}+E_{\rm vdW}^{\rm SI}\bigg).
\label{eqn:ediff}
\end{align}
Here $E_{\rm kin}^{{\rm Sol.}j}$ and $E_{\rm vdW}^{{\rm Sol.}j}$ are the kinetic and van der Waals energies of soliton $j$, while $E_{\rm kin}^{\rm I}$ and $E_{\rm vdW}^{{\rm Sol.}j}$ are those of the impurity. The van der Waals energy of the inter-species term is $E_{\rm vdW}^{\rm SI}$. In writing Eq.~\eqref{eqn:ediff}, we assume that those terms arising from the interaction of the tail of the impurity with that of the outer solitons (labelled Sol.1 and Sol.3) are negligible. Then, the energy difference $\delta E$ simplifies to
\begin{equation}
\delta E = E_{\rm kin}^{{\rm Sol.}1} + E_{\rm vdW}^{{\rm Sol.}1} + E_{\rm kin}^{{\rm Sol.}3} + E_{\rm vdW}^{{\rm Sol.}3},
\end{equation}
which demonstrates that $\delta E$ depends only on the outer solitons, and not the central soliton that carries the impurity. We can use the known analytical expression for the stationary bright soliton (Eq.~\eqref{eqn:1dsol}) profile to obtain an exact expression for $\delta E$. The total energy of each outer soliton is
\begin{equation}
E_{\rm sol} = -\frac{N_{1}^{3}mg_{11}^{2}}{24\hbar^2},
\end{equation}   
where $N_1$ and $g_{11}$ are the atom number and the quasi one-dimensional scattering parameter associated with the first component. To understand when thermal fluctuations play a role, we can form a dimensionless figure of merit as the ratio of the energy difference and the thermal energy present in the system at temperature $T$ as
\begin{equation}\label{eqn:rat}
\bigg|\frac{\delta E}{k_B T}\bigg|=\frac{N_{11}^{3}mg_{11}^{2}}{12\hbar^2 k_B T}.
\end{equation}
If this figure of merit satisfies $\delta E\gg k_BT$, then thermal effects should not play a dominant role in the dynamics of the soliton system. Likewise if $\delta E\ll k_BT$ then the trimer state will be destroyed by the thermal fluctuations. To gain insight into plausible experimental conditions for the observation of these states, we can again use the parameters of the experiment of Marchant \textit{et al.},\cite{marchant_2013}, where one has an $s$-wave scattering length $a_s=-11a_0$, an atom number $N_1=2000$, transverse oscillator strength $\omega_\perp=27$ Hz, with the atomic mass $m$ of $^{85}$Rb. Assuming an experimental temperature of $T\simeq1nK$, one obtains
\begin{equation}
\bigg|\frac{\delta E}{k_BT}\bigg|\simeq6.7,
\end{equation}
which satisfies the condition $\delta E/k_B T>1$. This rudimentary argument suggests that producing stable dynamics requires both a reasonable atom number as well as a low temperature. Since Eq.~\eqref{eqn:rat} depends on the cube of the atom number, it should in principle not be too difficult to satisfy this condition.} \me{Alternatively, one could also calculate the thermal stability of the trimer state from the quantity $E_{\rm tri}/k_BT$ alone. This could also give a deeper insight into the parameter regimes where this state is stable to thermal fluctuations and importantly how the inter-component scattering length $g_{\rm 12}$ affects this stability.}

\subsection{\label{sec:coh}Coherent Impurity Dynamics}

\noindent The dynamics of the impurity, presented in Fig.~\ref{fig:mol}(c) are suggestive that the soliton molecule could host coherent population dynamics. To investigate this effect we perform a comparison of the dynamics of the impurity component with a simple three level system, modeled in terms of a `vee' type atom. This model is chosen since the ground state energy (chemical potential) of the central soliton is slightly lower in energy; due to the presence of the impurity. Then, the equations of motion for the complex amplitudes $c_j(t)$ that determine the population of each soliton are 
\begin{equation}\label{eqn:vee}
\frac{d}{dt}\left(\begin{array}{c}c_1 \\ c_2 \\ c_3\end{array}\right)=-i\left(\begin{array}{ccc}0 & \sqrt{2}\Omega & 0 \\ \sqrt{2}\Omega & -\Delta & \sqrt{2}\Omega \\ 0 & \sqrt{2}\Omega & 0\end{array}\right)\left(\begin{array}{c}c_1 \\ c_2 \\ c_3\end{array}\right).
\end{equation}
We can connect the solutions $c_j(t)$ of Eqs.~\eqref{eqn:vee} to the populations presented in Fig.~\ref{fig:mol} since $P_{\rm Sol.\#j}(t)=|c_j(t)|^2$. The dynamical system described by Eqs.~\eqref{eqn:vee} introduce the `Rabi' frequency $\Omega$, which defines the frequency of population transfer between solitons, and the effective `detuning' $\Delta$. The total population is a conserved quantity given by $\sum_j |c_j(t)|^2=N_2$.
\begin{figure}[t]
\includegraphics[width=\columnwidth]{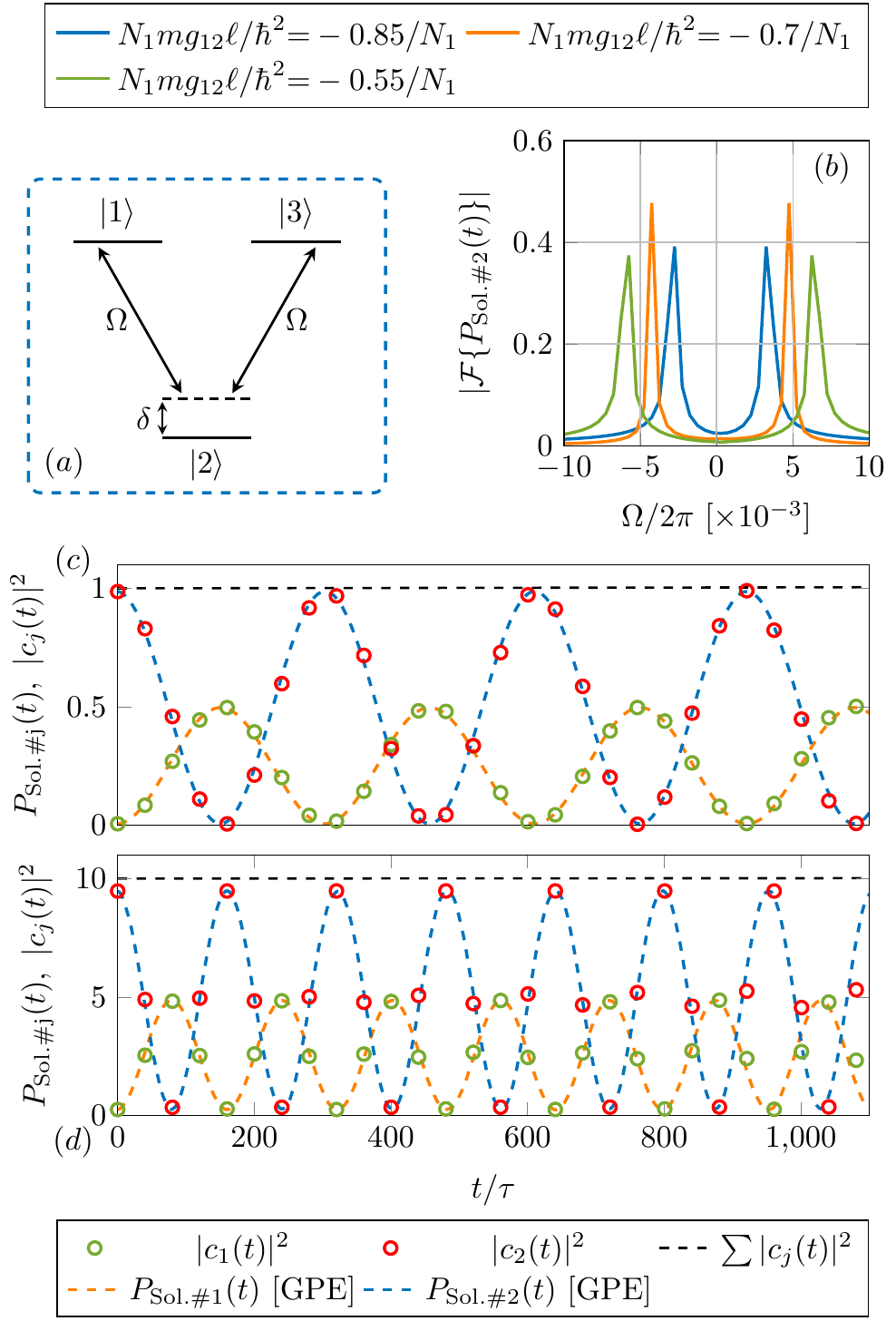}
\caption{\label{fig:rabi}Rabi model comparison. (a) shows the analogous level scheme for the three level atom. (b) shows examples of the Fourier transform of the impurity populations obtained numerically from the GPE for various values of $\me{g_{\rm 12}}$. Panels (c) and (d) show comparisons of the solutions to the Rabi model \mje{(see Eq.~\eqref{eqn:vee} and Eqs.~\eqref{eqn:vsol})} with GPE simulations.}
\end{figure}

Figure \ref{fig:rabi} shows comparisons of the Rabi model, Eq.~\eqref{eqn:vee} with Gross-Pitaevskii simulations. The analogous `vee' atom level diagram is shown in Fig.~\ref{fig:rabi} (a), where the states $|1\rangle,|2\rangle$ and $|3\rangle$ represent the potential generated by the left, middle and right soliton felt by the impurity atoms. Then one can associate a state vector with Eq.~\eqref{eqn:vee} for the impurity of the form
\begin{equation}
|\psi_{\rm imp}\rangle=\sum_{j=1}^{3}c_{j}(t)|j\rangle.
\end{equation}
\mje{Due to the simplicity of the effective model Eq.~\eqref{eqn:vee}, we can obtain exact expressions for the time-dependent amplitudes $c_j(t)$ using the eigenbasis of the Hamiltonian matrix appearing on the right-hand-side of Eq.~\eqref{eqn:vee}. The three orthogonal eigenvectors of this system are $\nu_0=(-1,0,1)$ and $\nu_{\mp}=(1,i\lambda_{\mp}/\sqrt{2}\Omega,1)$ with the associated eigenfrequencies $\lambda_0=0$ and $\lambda_{\mp}=\frac{i}{2}(\Delta\mp\Omega_{\rm d})$ where $\Omega_{\rm d}=\sqrt{16\Omega^2+\Delta^2}$. Using the initial conditions $c_{1,3}(t{=}0){=}0$ and $c_2(t{=}0){=}\sqrt{N_2}$ the solutions to Eq.~\eqref{eqn:vee} can be written      
\begin{subequations}\label{eqn:vsol}
\begin{align}
c_{1,3}(t)&=\sqrt{N_2}\frac{\sqrt{2}\Omega}{\Omega_{\rm d}}e^{i\frac{\Delta t}{2}}2i\sin\bigg(\frac{\Omega_{\rm d}t}{2}\bigg),\\
c_{2}(t)=&\sqrt{N_2}e^{i\frac{\Delta t}{2}}\bigg[\cos\bigg(\frac{\Omega_{\rm d}t}{2}\bigg){+}i\frac{\Delta}{\Omega_{\rm d}}\sin\bigg(\frac{\Omega_{\rm d}t}{2}\bigg)\bigg].
\end{align}
\end{subequations}
The solutions given by Eqs.~\eqref{eqn:vsol} can be used to gain insight into the nature of the underlying tunneling effect responsible for the impurities transport inside the solitons. To do this, we calculate the tunneling current $\mathcal{J}_{\rm t}(t)=-i[c_{1,3}(t)c_{2}^{*}(t) - c_{2}(t)c_{1,3}^{*}(t)]$ using the solutions for $c_{j}(t)$ from Eqs.\eqref{eqn:vsol} giving
\begin{equation}\label{eqn:jc}
\mathcal{J}_{\rm t}(t)=\frac{\sqrt{8}N_2\Omega}{\Omega_{\rm d}}\sin(\Omega_d t),
\end{equation}
which shows that the tunneling current $\mathcal{J}_{\rm t}(t)$ attains a maximum or minimum value when $\Omega_{\rm d}t=\frac{1}{2}(2n+1)\pi$ and $n$ is zero or a positive integer, which as can be seen from Fig.~\eqref{fig:rabi} (c) and (d) is exactly when the impurity in the inner soliton (soliton two) has a maximum in its impurities population. Likewise, the tunneling current $\mathcal{J}_{\rm t}(t)$ goes to zero when $\Omega_{\rm d}t=n\pi$ for even integer $n$; which corresponds to when the outer solitons (soliton one and two) have their maximum impurity population.  
}

To \mje{numerically} obtain the Rabi frequency $\Omega_{\rm d}$, we take the Fourier transform $\mathcal{F}\{P_{\rm Sol.\#2}(t)\}$ which is shown in Fig.~\ref{fig:rabi}(b), for $N_{1}^{2}m\me{g_{\rm 12}}\ell/\hbar^2=-0.85,-0.7,-0.55$. Here the system parameters are $N_1m\me{g_{\rm 12}}\ell/\hbar^2=-8/N_1$, with $N_1=1,000$. Then for the examples (c) and (d) one has $N_2=1$ and $N_2=10$ impurity atoms respectively. The outer solitons are placed at $x_0=\pm6\ell$, the initial phase of the central soliton was $\delta=\pi$, and the detuning is $\Delta=5\times10^{-3}\tau^{-1}$. Each inter-component scattering length gives a single peaked spectrum\me{, shown in Fig.~\ref{fig:rabi}(b).} The lower panels (c) and (d) of Fig.~\ref{fig:rabi} show comparisons between the solutions $c_j(t)$ obtained from Eq.~\eqref{eqn:vee} and the impurity populations calculated from the GPE via Eq.~\eqref{eqn:popi}. In both presented examples (c) and (d), the dashed lines represent the impurity populations computed from GPE simulations, while the circles are the Rabi model data. The dashed black line shows the total population $\sum_j|c_j(t)|^2=N_2$. In both presented cases, Fig.~\ref{fig:rabi} (c) and (d) we find excellent agreement to the Rabi model. It is important to note that at much longer times, the outer solitons are attracted towards the central soliton, which causes the effective Rabi frequency $\Omega_{\rm d}$ to increase, but by sensibly choosing the system parameters such that the outer solitons are not initially too close to the central soliton, good agreement to Eq.~\eqref{eqn:vee} is obtained. \mje{The coherent oscillations presented in Fig.~\ref{fig:rabi} could form the basis for future applications. In particular, the identification of these types of dynamics could find practical application in atomtronics \cite{seaman_2007,amico_2017} and quantum information processing \cite{kok_2010}, where the coherent dynamics of atomic systems are a required ingredient for many effects of interest in these fields.}   

\subsection{\label{sec:ipr}Impurity Localization Transition}
\noindent The dynamics of the impurity presented in Fig.~\ref{fig:mol} and \ref{fig:rabi} are suggestive of rich transport behavior. To understand the transport properties of the multi-soliton system further, we probe the dynamics of the three soliton system across the full parameter space. Obtaining a well-behaved, intuitive measure for the multiple soliton system is challenging. To understand the effect of the various system parameters on the dynamics of the impurity, we employ the inverse participation ratio (IPR) as a measure to quantify the dynamics of the system. The inverse participation ratio provides a well-behaved measure of how localized a particular state is \cite{wegner_1980}. \mje{Related to this, recent work has also examined the  effect of `dynamical localization' in dynamical optical lattice potentials \cite{major_2018}.} In particular we wish to calculate
\begin{equation}\label{eqn:ipr}
\frac{1}{\mathcal{P}(t)}=\frac{\int dx |\psi_{2}(x,t)|^4}{(\int dx |\psi_{2}(x,t)|^2)^2}.
\end{equation}
\begin{figure}[t]
\includegraphics[width=\columnwidth]{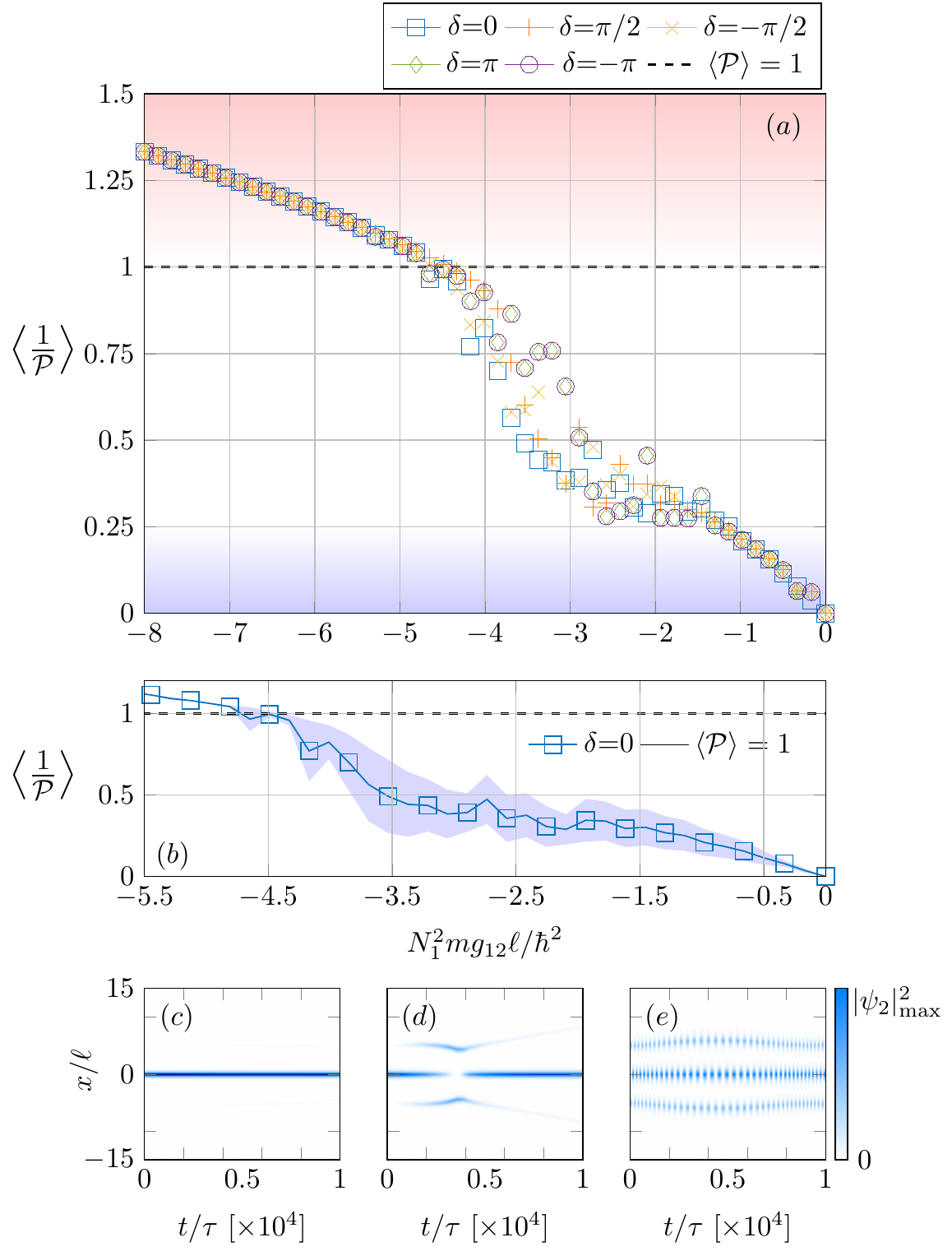}
\caption{\label{fig:ipr}The inverse participation ratio (IPR) calculated from Eqs.~\ref{eqn:ipr} and \ref{eqn:avg_ipr}. (a) shows the IPR for different initial phases, $\delta$. (b) shows the IPR (solid blue squares) and it's fluctuations (light shaded blue) in terms of the standard deviation for $\delta=0$. The bottom row of panels shows individual simulations for $N_{1}^{2}m\me{g_{\rm 12}}\ell/\hbar^2=-3.85,-2.9,-0.98$ for (c), (d) and (e) respectively. The dashed black line in (a) and (b) separates localized states at $\langle\mathcal{P}\rangle=1$.}
\end{figure}
For non-interacting spatially localized states, the inverse participation ratio takes a value of one such that $\mathcal{P}(t)^{-1}=1$, while delocalized states are found instead when $\mathcal{P}(t)^{-1}\ll1$. \mje{This definition is however strictly speaking only applicable to non-interacting systems, the introduction of mean-field interactions can yield value of the IPR that are greater than one. Nonetheless this quantity still provides a useful measure of the impurities spatial dynamics.} Since we are dealing with a two component system where both components evolve dynamically, it is necessary to consider the {\it time averaged} version of Eq.~\eqref{eqn:ipr} in order to make a meaningful analysis. The time average of Eq.~\eqref{eqn:ipr} is defined as
\begin{equation}\label{eqn:avg_ipr}
\left\langle\frac{1}{\mathcal{P}}\right\rangle=\frac{1}{T}\int_{0}^{T}\frac{dt}{\mathcal{P}(t)}
\end{equation}
where $T$ defines the length of the particular numerical simulation. To investigate the behavior of the IPR, Eq.~\eqref{eqn:ipr}, and in particular its time average given by Eq.~\eqref{eqn:avg_ipr}, we perform numerical simulations using the initial state defined by Eq.~\eqref{eqn:itri}, with the parameters $N_1=1,000$ and $N_2=1$ giving the atom numbers for the soliton and impurity respectively, while $N_1m\me{g_{\rm 11}}\ell/\hbar^2=-8/N_1$ defines the strength of the van der Waals parameters for the first component. The outer solitons were placed at $x_0=\pm5\ell$ from the origin. Finally, each individual simulation was run for $T/\tau=10^{4}$ units of time. Long time simulations of the trimer state are presented in figure \ref{fig:ipr}. The time-averaged inverse participation ratio is shown as a function of the dimensionless inter-component scattering parameter $N_{1}^{2}m\me{g_{\rm 12}}\ell/\hbar^2$ in Fig.~\ref{fig:ipr}(a). Here, data is presented for several different initial phase differences: $\delta=0,\pm\pi/2,\pm\pi$. The dynamics can be divided into three regions, a localized region (red gradient), a delocalized region, (blue gradient) and an intermediate region (white). The trend in (a) shows that for large negative $\me{g_{\rm 12}}$ the impurity is localized, since $\mathcal{P}\geq1$, here all data fall onto a common curve. \mje{The observed behaviour of the IPR attaining values greater than one notably differs from its original definition where localized states are defined for $\langle\mathcal{P}(t)\rangle=1$ only. We attribute this departure to the fact that we are considering an interacting, rather than non-interacting system. Then as} the scattering length is increased, the impurity starts to delocalize across the three solitons and individual datum no longer follow a common trend, instead the particular value of $\langle\mathcal{P}^{-1}\rangle$ one obtains is found to be sensitive to the initial phase $\delta$. As the scattering parameter $\me{g_{\rm 12}}$ approaches zero, the data again fall onto a common curve, and the impurity is completely delocalized between the three solitons. In this region stable molecules are found that support this effect.

To understand the impurity dynamics in the intermediate region, (white region in Fig.~\ref{fig:ipr}) the fluctuations during dynamics of the IPR (Eq.~\eqref{eqn:ipr}) are studied by calculating the standard deviation in Fig.~\ref{fig:ipr}(b). The standard deviation of the IPR is plotted with the average of the IPR (light blue shading and solid blue respectively). One can see that the fluctuations associated with Eq.~\eqref{eqn:avg_ipr} start to grow as $\langle\mathcal{P}^{-1}\rangle$ falls below one. \mje{Indeed, it would seem within the mean-field model considered in this work one can attribute the point $\langle\mathcal{P}(t)\rangle=1$ as the point in the parameter space where fluctuations of the IPR grow from zero and the impurity begins to delocalize between the outer solitons.} The final row of figures shown in Fig.~\ref{fig:ipr} shows example dynamics for each dynamical region. In particular Fig.~\ref{fig:ipr}(c) shows an example of a localized impurity. Then Fig.~\ref{fig:ipr}(d) shows an example of the intermediate regime, and finally Fig.~\ref{fig:ipr}(e) shows the delocalized region.   

\section{\label{sec:sum}Conclusions}

\noindent In this work we investigated the scattering properties of a two-component Bose condensate with wholly attractive mean field interactions. By interpreting the second component as an impurity, this system was found to support unusual transport phenomena, including the appearance of a dimer like phase close to zero inter-component scattering length, where a pair of bright solitons in the first component can coherently transfer the impurity between each other many times. Such an effect could be useful for example in the emergent field of atomtronics, where atomic systems are used to build circuits analogous to their electronic counterparts. The ability to use solitary waves to coherently shuttle atomic density over macroscopic distances could form a novel tool in this endeavor. 

It was also found that stable soliton molecules formed from three solitons can also be produced in parameter regimes where the equivalent single component system is unstable to the formation of molecular bound states. This stability was attributed to the nontrivial phase winding that occurs during dynamical evolution of the two-component system. Since the impurity that constitutes the second component can effectively delocalize itself across the whole system, the atom number of both components of the gas can change. Accompanying this change is a winding of the phase, which for a critical scattering length can be favorable to the formation of three soliton molecules. The population dynamics of the impurity was scrutinized using a simple three level atomic `Rabi' model. For sensible choices of parameters excellent agreement was obtained with GPE simulations. Finally, the trimer-impurity system was analyzed using the tools of localization theory. It was found that the impurity undergoes a delocalization as a function of the inter-component scattering length.

It would be interesting to investigate the effect of trapping fermions in this physical setup, in a similar spirit to the experiment of Ref.~\cite{desalvo_2017}. The ability to build larger systems of solitons with this particular system opens a novel avenue in studying lattices formed from solitary waves, with the twist that one can have different numbers of impurities present, which could be used to study effects analogous to condensed matter, for example a soliton-Hubbard model could be potentially explored, as well as understanding the generalized Toda lattice that this system would constitute.

\section{Acknowledgements}

\noindent We thank Robert Dingwall and Tom Billam for useful discussions. MJE acknowledges support as an Overseas researcher under Postdoctoral Fellowship of Japan Society for the Promotion of Science. This work was supported by the Okinawa Institute of Science and Technology Graduate University.

\end{document}